\documentclass[conference,a4paper]{IEEEtran}
\usepackage{cite}
\ifCLASSINFOpdf
\usepackage[pdftex]{graphicx}
\graphicspath{./figures}
\DeclareGraphicsExtensions{.pdf,.jpeg,.png}
\else
\fi
\usepackage[cmex10]{amsmath}
\usepackage{array}
\usepackage{url}
\usepackage[utf8]{inputenc}  
\usepackage{citesort}
\usepackage[T1]{fontenc} 
\usepackage{mathtools}
\usepackage{amsfonts}
\usepackage{amssymb}
\usepackage[english]{babel}
\usepackage{algorithm}
\usepackage[noend]{algpseudocode}
\usepackage{float}
\usepackage{multirow}
\usepackage{bbm}
\usepackage{color}

\newcommand\blfootnote[1]{%
  \begingroup
  \renewcommand\thefootnote{}\footnote{#1}%
  \addtocounter{footnote}{-1}%
  \endgroup
}

\DeclareUnicodeCharacter{00A0}{~}

\def \({\left(}
\def \){\right)}
\def \[{\left[}
\def \]{\right]}

\newcommand{\tbf}[1]{{\textbf{#1}}}

\newcommand{\br}{{\textbf {r}}}
\newcommand{\bu}{{\textbf {u}}}
\newcommand{\bF}{{\textbf {F}}}

\newcommand{\bx}{{\textbf {x}}}

\newcommand{\by}{{\textbf {y}}}

\newcommand{\bz}{{\textbf {z}}}

\newcommand{\bs}{{\textbf {s}}}

\newcommand{\be}{\begin{equation}}
\newcommand{\ee}{\end{equation}}
\newcommand{\bea}{\begin{eqnarray}}
\newcommand{\eea}{\end{eqnarray}}

\begin{document}

\title{A Compressed Sensing Approach for\\ Distribution Matching}
\title{A Compressed Sensing Approach for\\ Distribution Matching}

\author{
    Mohamad Dia\\
    LTHC - EPFL, Lausanne, Switzerland \\
    mohamad.dia@epfl.ch
  \and
    Vahid Aref and Laurent Schmalen\\
    Nokia - Bell Labs, Stuttgart, Germany\\
    \{vahid.aref, laurent.schmalen\}@nokia-bell-labs.com
}

%
\maketitle
\blfootnote{M. Dia acknowledges funding from SNSF (grant no. 200021-156672).}

\begin{abstract}
In this work, we formulate the fixed-length distribution matching as a Bayesian inference problem. Our proposed solution is inspired from the compressed sensing paradigm and the sparse superposition (SS) codes.  First, we introduce sparsity in the binary source via position modulation (PM). We then present a simple and exact matcher based on Gaussian signal quantization. At the receiver, the dematcher exploits the sparsity in the source and performs low-complexity dematching based on generalized approximate message-passing (GAMP). We show that GAMP dematcher and spatial coupling lead to an asymptotically optimal performance, in the sense that the rate tends to the entropy of the target distribution with vanishing reconstruction error in a proper limit. Furthermore, we assess the performance of the dematcher on practical Hadamard-based operators. A remarkable inherent feature of our proposed solution is the possibility to: $i)$ perform matching at the symbol level (nonbinary); $ii)$ perform joint channel coding and matching. 
\end{abstract}

\section{Introduction}

Distribution matching has recently attracted lots of attention in long-haul fiber optical communications. As an inverse of data compression, a distribution matcher maps a discrete memoryless source, namely i.i.d. Bernoulli$(1/2)$ bits, into a sequence of symbols distributed according to a target distribution. A dematcher is required to perform the inverse operation and recover the original source with a certain reliability.

As a primary application, distribution matching is used for probabilistic shaping
\cite{Forney_84,Steiner_Bocherer_16_comparison}
in order to imitate the capacity achieving input distribution of the channel and increase the spectral efficiency. The distribution matching task in probabilistic shaping can be done at the bit level by introducing bias in the binary source followed by a high-order modulation scheme that yields nonuiform symbols. However, one can perform distribution matching at the symbol level by directly mapping the binary sequence into the desired symbols (e.g. nonunifrom QAM symbols). Distribution matching is also used to achieve the capacity of asymmetric channels \cite{Mondelli_Urbanke_Hassani_14_assymetricChannels} and for rate adaptation \cite{MacKay_99_good-error-correcting}.

Optimal variable-length distribution matching schemes with offline algorithms were proposed in \cite{Kschischang_Pasupathy_93_optimal-nonuniform,Ungerboeck_02_huffman-shaping,Bocherer_11_dyadic,Amjad_Bocherer_13_fixed-to-variable}. A low-complexity online algorithm based on arithmetic coding was introduced in \cite{Cai_Ho_Yeung_07_probabilistic-capacity,Baur_Bocherer_15_arithmetic}. Variable-length schemes require large buffer sizes and suffer from error propagation and synchronization problems \cite{Kschischang_Pasupathy_93_optimal-nonuniform}. Recently, an asymptotically optimal fixed-length and low-complexity distribution matcher was introduced in \cite{Schulte_Bocherer_16_constant}.

All the aforementioned schemes are lossless. However, their practical implementations require a separate forward error correction code to be added on top of the distribution matcher \cite{Steiner_Bocherer_16_comparison}, which might incur a rate loss and error propagation for finite blocklengths.
In this work,  we propose a scheme which inherently performs joint channel coding and distribution matching. In particular, we formulate the fixed-length distribution matching as a Bayesian inference problem. The formulation is inspired from the compressed sensing paradiagm \cite{Donoho_06_compressed-sensing,Candes_Tao_06_near-optimal} and sparse superposition (SS) codes \cite{Barron_Joseph_10_toward-fast-reliable,Barbier_17_capacity-achieving,Rush_17,Dia_16_ISIT}.
Moreover, we provide a low-complexity
algorithm based on generalized approximate message-passing (GAMP) \cite{Donoho_09,Rangan_12_gamp} and spatial coupling. The proposed scheme is asymptotically optimal and it is motivated by the recent success of GAMP in quantized compressed sensing \cite{QunatizedCS_2012_Rangan} and SS codes \cite{Dia_16_ITW,Dia_17_ISIT,Dia_17_transactions}.

For the proposed scheme, the algorithmic performance under GAMP dematcher and the Bayes-optimal performance, under optimal dematcher, can be tracked by the state evolution (SE) and potential function. 
We show via SE analysis and numerical simulations that GAMP operates up to an ``algorithmic rate'' with a nonnegligible gap to the information theoretical rate. However, we illustrate that the GAMP dematcher on a spatially coupled version of the problem is asymptotically optimal in the blocklength, in the sense that the algorithmic rate saturates the Bayes-optimal performance which, in turn, tends to the entropy of the target distribution in a proper limit. Furthermore, unlike the existing approaches, the target distribution is attained for all blocklengths due to the simplicity of the matcher which is based on quantizing a Gaussian signal.

Bearing in mind practical implementations, we assess the performance of the dematcher on Hadmard-based operators that allow for substantial decrease in the complexity and memory needs. 
It is noteworthy to mention that our approach provides a single-shot solution by performing distribution matching at the symbol level in addition to the possibility of implementing joint channel coding on memoryless channels.

\section{Distribution Matching}
In binary distribution matching, one is ideally interested in mapping a binary sequence $\bu \in \{0,1\}^{m}$ with i.i.d. Bernoulli$(1/2)$ bits into another
discrete sequence $\by \in \mathcal{A}^{M}$ having a target marginal distribution $P_{Y}$. The mapping is done such that $\bu$ is perfectly reconstructed from $\by$. We call $\bu$ the \emph{source} and $\by$ the \emph{target}. Let $Y$ be the target random variable with alphabet $\mathcal{A}$ and distribution $P_{Y}$. The maximal achievable rate (or the information theoretical rate) of lossless distribution matching is given by
\begin{align}\label{eq:achievableRate}
R = \frac{m}{M} \le H(Y),
\end{align}
where $H(Y)$ is the entropy of $Y$. In the binary-to-binary case, $\bu$ is mapped to another binary sequence $\by \in \{0,1\}^{M}$ with $M$ Bernoulli$(p^{\star})$ bits, where $p^{\star}$ represents the target distribution. 
%


Note that one can frame this problem as the inverse of the lossless source coding. Consequently, a natural approach to solve the distribution matching problem is to use variable-length prefix-free source coding schemes such as Huffman codes \cite{Kschischang_Pasupathy_93_optimal-nonuniform,Ungerboeck_02_huffman-shaping,Bocherer_11_dyadic,Amjad_Bocherer_13_fixed-to-variable} or low-complexity arithmetic codes \cite{Cai_Ho_Yeung_07_probabilistic-capacity,Baur_Bocherer_15_arithmetic}. In this case, perfect reconstruction is guaranteed for all blocklengths, while the distortion measure is defined to be the normalized Kullback-Leibler (KL) divergence between the matcher distribution and the target distribution \cite{Schulte_Bocherer_16_constant}. As the blocklength increases, the rate of the aforementioned schemes tends to the maximal achievable rate \eqref{eq:achievableRate} with vanishing normalized KL divergence between the matcher and target distributions. However, the main limitation of these schemes is the varying transmission rate. Recently, a fixed-rate approach based on constant composition arithmetic coding was introduced in \cite{Schulte_Bocherer_16_constant}.

Another approach to solve the distribution matching is to employ a forward error correcting code \cite{Mondelli_Urbanke_Hassani_14_assymetricChannels} as done in lossless source coding
\cite{Caire_03_lossless,Cronie_10_losslessPolar}. In this approach, the target distribution can be matched while the distortion measure is the reconstruction error, between the original source and the reconstructed one, that vanishes in the blocklength. Although this approach might be erroneous at finite blocklengths, it remains very useful for many application scenarios because of its amenability to perform joint channel coding and matching. Following this second approach, we propose a solution that employs the SS codes for distribution matching and relies on the recent success of GAMP algorithm for such codes \cite{Dia_16_ITW,Dia_17_ISIT,Dia_17_transactions}.

\section{Compressed Sensing Approach}
Our proposed solution, depicted in Fig. \ref{fig:blockDiag}, employs the SS codes used for general channel \cite{Dia_17_transactions} to perform distribution matching. 
One can formulate the distribution matching as a SS code on a deterministic nonlinear channel, and hence leverage the GAMP algorithm to perform the dematching. The GAMP dematcher identifies an \emph{effective channel}, which can be a concatenation of the deterministic matcher (quantizer) with a noisy channel, over which the estimation is performed. Thus, distribution matching and channel coding can be jointly performed. In this work, we focus on the distribution matching part for symplicity. Hence, our channel is a quantization of a Gaussian signal that yields the target distribution. Note that in \cite{Dia_16_ITW,Dia_17_transactions}, SS codes were used for forward error correction over general channel while other techniques \cite{Mondelli_Urbanke_Hassani_14_assymetricChannels} were proposed to perform the distribution matching task. The main contribution of this work is to show that SS codes can be used to perform distribution matching concurrently for any discrete alphabet.

\subsection{Matcher}
\begin{figure}[!t]
\centering
\includegraphics[draft=false,width=.34 \textwidth, height = 100pt]{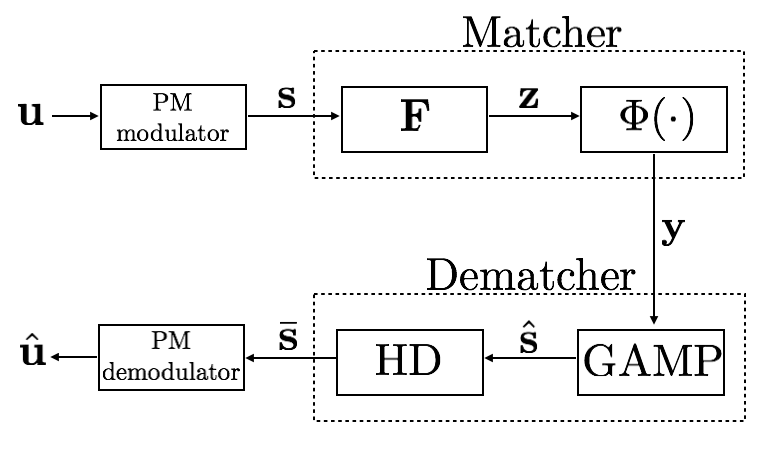}
\caption{
The original source $\tbf u$ is mapped to a sparse signal $\bs$ through a PM modulator. A quantizer $\Phi(\cdot)$ is then applied to obtain the target sequence $\by$. The GAMP algorithm provides soft valued estimate $\hat{\bs}$ of $\bs$ in the MMSE sense. A simple hard decision (HD) scheme is used to provide the binary decoded message $\bar{\bs}$ by setting the most biased component in each section of $\hat{\bs}$ to $1$ and the others to $0$. The reconstructed version $\hat{\tbf{u}}$ of the original source $\tbf u$ can be easily recovered from $\bar{\bs}$ using PM demodulator.}
\label{fig:blockDiag}
\end{figure}

In order to use the compressed sensing and AMP paradigms, we need to introduce sparsity in $\bu$. This can be done via simple position modulation (PM) scheme (see Fig. \ref{fig:blockDiag}). Take $m=L\log_2(B)$, with $B$ chosen to be a power of 2. The original source can be seen as a vector made of $L$ \emph{sections}, $\bu = [\bu_1, \dots, \bu_L]$, where each section $\bu_l$, $l\in\{1, \dots, L\}$ is a $\log_2(B)$-dimensional vector. We call $B$ the \emph{section size}. The original source is then mapped to a sparse \emph{signal} $\bs$ made of $L$ sections. Each section $\bs_l$ is a $B$-dimensional vector with a single nonzero component that is equal to $1$. The position of the non-zero component in $\bs_l$ is 
specified by the binary representation of $\bu_l$. For example if $B=4$ and $L=5$, a 
valid source is $\bu = [00,01,11,10,01]$ which corresponds to $\bs = [0001,0010,1000,0100,0010]$. We set $N=LB$.

A fixed \emph{coding matrix} $\bF\in \mathbb{R}^{M \times N}$ is taken with i.i.d real Gaussian entries distributed as $\mathcal{N}(0, 1/L)$. We use this matrix to obtain a \emph{codeword} $\bz=\bF\bs\in \mathbb{R}^{M}$ with i.i.d. standard Gaussian entries. The matching task consists of quantizing the Gaussian codeword entries through a quantizer $\Phi(\cdot)$ acting componentwise with
\begin{align}\label{eq:quantizer}
y_i = \Phi(z_i) = \Phi([\bF\bs]_i), \quad i = 1,\dots, M,
\end{align}
such that the output is distributed according to a given target distribution. Note that one can look at $\Phi(\cdot)$ as a deterministic nonlinear channel leading to the target distribution.

For a general $q$-ary discrete target distributions (e.g. PAM or QAM symbols), the quantizer $\Phi$ uses biased $q$-quantiles of the Gaussian distribution for quantization. Specifically, let $Y$ be a discrete random variable with $q$-ary alphabet $\mathcal{A} = \{a_1, \dots, a_q \}$ and distribution $P_{Y}(a_k) = P_k$ ($k \in \{1,\dots,q\}$). The quantizer is defined by
\begin{align}\label{eq:quantizer_qary}
\Phi(z) = a_k \quad \text{if} \, z \in ]c_{k-1},c_k],
\end{align}
with
\begin{align}\label{eq:quantizer_qary_1}
c_k = 
\begin{cases}
- \infty \qquad \qquad \qquad \quad \, \, k =0,\\
Q^{-1}(1-\sum_{j=1}^{k}P_j) \quad  k = 1,\dots, q,
\end{cases}
\end{align}
where $Q^{-1}(\cdot)$ is the inverse of the Gaussian $Q$-function.

\begin{figure}[!t]
\centering
\includegraphics[draft=false,width=.34\textwidth, height = 122pt]{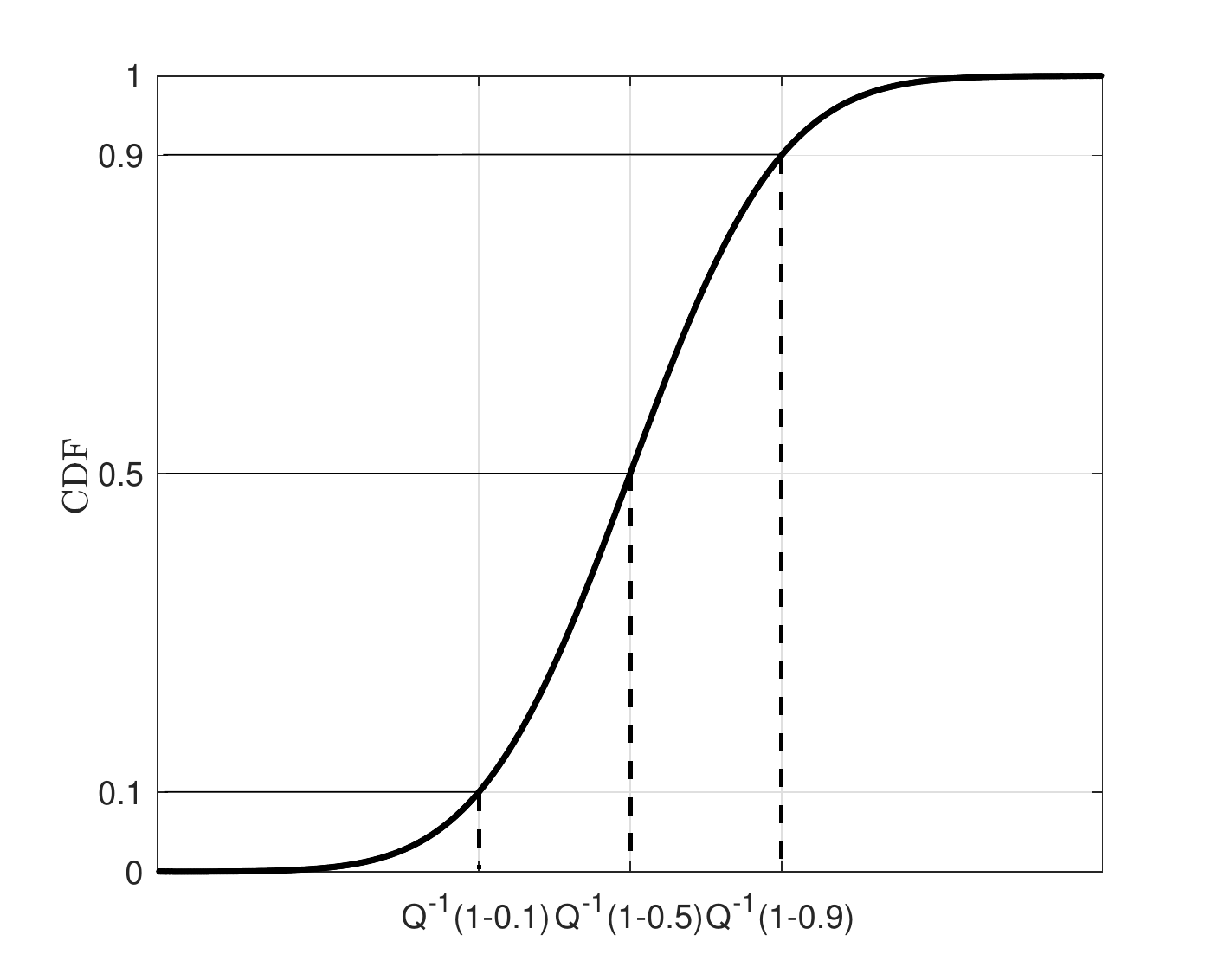}
\caption{Quantization intervals of a standard Gaussian signal to nonuniform $4$-PAM constellation $\mathcal{A} = \{a_1, a_2, a_3,a_4\}$ with $P_{Y} = [0.1, 0.4, 0.4,0.1] $.
}
\label{fig:quantizer_qary}
\end{figure}

Note that this simple matching operation ensures that the target distribution is attained for all blocklengths. The quantization intervals of a nonuniform $4$-PAM constellation based on the cumulative distribution function of a standard Gaussian are shown in Fig. \ref{fig:quantizer_qary}.
\subsection{Dematcher}
The dematching task is to recover a sparse signal $\bs$, and hence $\bu$, from quantized random projections $\by$ as depicted in Fig. \ref{fig:blockDiag}. The sparsity introduced in the signal by PM can be harnessed at the dematcher in a compressed sensing fashion. Namely, the dematching can be interpreted as a compressed sensing problem with structured sparsity.
Consequently, the same algorithms and analysis tools used in compressed 
sensing theory, such as GAMP algorithm and SE, can be used for the dematching task as done for SS codes \cite{Barbier_17_capacity-achieving,Dia_17_transactions}. For a Bernoulli$(1/2)$ source and PM scheme, the sections of $\bs$ are uniformly distributed over all the possible $B$-dimensional vectors with a single nonzero component that is equal to 1. The prior of each section is denoted by $p_0(\bs_l)$.

In a Bayesian setting, the estimation of the signal $\bs$, based on the observed target $\by$ and a fixed matrix $\bF$, can be done in a minimum mean-square error (MMSE) sense or maximum a-posteriori (MAP) sense. This necessitates the computation of the posterior distribution of $\bs$ given $\by$ and $\bF$ on a dense graphical model.
Therefore, one can use an iterative message-passing algorithm such as GAMP, which was first introduced in compressed sensing \cite{Donoho_09,Rangan_12_gamp} and then adapted to account for any structured $B$-dimensional prior distribution \cite{Barbier_17_capacity-achieving,Rush_17,Dia_17_transactions}. 
The GAMP algorithm uses Gaussian and quadratic approximations that yield a sequence 
of disjoint estimation problems under an \emph{equivalent Gaussian noise}.
The real physical channel appears in the computation of the moments of the equivalent Gaussian noise. The physical channel in our distribution matching problem is a deterministic highly nonlinear channel defined by the quantizer \eqref{eq:quantizer}. Thus, the GAMP algorithm of \cite{Dia_17_ISIT,Dia_17_transactions} can be adapted to act on such channel.

Algorithm \ref{alg:gamp} shows the steps of GAMP. This algorithm was first introduced in \cite{Rangan_12_gamp} and then adapted to the structured prior of SS codes in \cite{Dia_17_ISIT,Dia_17_transactions}. The ``$\circ$'' symbol in Algorithm \ref{alg:gamp} indicates that the corresponding operator is taken componentwise (componentwise multiplication, square, inverse). 
\begin{algorithm}[H]
\caption{GAMP ($\by,\bF,B,t_{\rm max}$)}\label{alg:gamp}
\begin{algorithmic}[1]
\State $\hat{\bs}^{(0)} \quad \gets \, \, \mathbf{0}_{N,1}$
\State ${\boldsymbol{\sigma}}^{(0)} \,\,\, \,\gets \, \, (1/B) \mathbf{1}_{N,1}$
\State $\bx^{(-1)} \, \,\gets \, \, \mathbf{0}_{M,1}$
\State $t \, \qquad \gets \, \, 0$
\While{$t< t_{\rm max}$}
\State ${\boldsymbol{\eta}}^{(t)} \quad \gets \, \, \textbf{F}^{\circ2}{\boldsymbol{\sigma}}^{(t)}$
\State ${\textbf{p}}^{(t)} \quad \,\gets \, \, \textbf{F}\hat{\bs}^{(t)} - {\boldsymbol{\eta}}^{(t)} \circ \bx^{(t-1)}$
\State $\bx^{(t)} \quad \, \gets \, \, g_\text{out}(\textbf{p}^{(t)}, \by, {\boldsymbol{\eta}}^{(t)})$ \label{step:gout}
\State ${\boldsymbol{\zeta}}^{(t)} \quad \, \gets \, \, f_\text{out}(\textbf{p}^{(t)}, \by, {\boldsymbol{\eta}}^{(t)})$ \label{step:fout}
\State ${\boldsymbol{\tau}}^{(t)} \quad  \gets \, \, {\big[\big(( {\boldsymbol{\zeta}}^{(t)})^{\intercal}\textbf{F}^{\circ2}\big)^{\intercal}\big]}^{\circ-1}$
\State $\textbf{r}^{(t)} \quad \, \, \gets \, \, \hat{\bs}^{(t)} + {\boldsymbol{\tau}}^{(t)} \circ \big((\bx^{(t)})^{\intercal}\textbf{F}\big)^{\intercal}$
\State $\hat{\bs}^{(t+1)} \, \, \gets \, \, g_\text{in}(\textbf{r}^{(t)}, {\boldsymbol{\tau}}^{(t)})$ \label{step:gin}
\State ${\boldsymbol{\sigma}}^{(t+1)} \gets \, \, \hat{\bs}^{(t+1)} - \big(\hat{\bs}^{(t+1)}\big)^{\circ2}$ \label{step:gin_p}
\State $t \qquad \, \, \gets \, \, t+1$
\EndWhile\label{gampwhile}
\State \textbf{return} $\hat{\bs}^{(t)}$
\end{algorithmic}
\end{algorithm}
Steps \ref{step:gout} and \ref{step:fout} of Algorithm \ref{alg:gamp} depend on the the quantizer \eqref{eq:quantizer}. For the $q$-ary quantizer given in \eqref{eq:quantizer_qary} and \eqref{eq:quantizer_qary_1}, the $i^{th}$ entry of $g_\text{out}$ and  $f_\text{out}$ take the following forms respectively
\begin{align*}
[g_{\text{out}}(\textbf{p}, \by, {\boldsymbol{\eta}})]_i &=
\frac{\sum_{k=1}^{q} \delta(y_i \!-\! a_k) \big(Q^{'}_{k\!-\!1}(p_i,\eta_i) \!-\!Q^{'}_{k}(p_i,\eta_i)\big)}{\sum_{k=1}^{q} \delta(y_i\!-\!a_k) \big(Q_{k\!-\!1}(p_i,\eta_i) \!-\!Q_{k}(p_i,\eta_i)\big)}, \nonumber \\
[f_{\text{out}}(\textbf{p}, \by, {\boldsymbol{\eta}})]_i &=
\big([g_{\text{out}}(\textbf{p}, \by, {\boldsymbol{\eta}})]_i\big)^2 \nonumber \\
 &-\frac{\sum_{k=1}^{q} \delta(y_i\!-\!a_k) \big(Q^{''}_{k\!-\!1}(p_i,\eta_i) \!-\! Q^{''}_{k}(p_i,\eta_i)\big)}{\sum_{k=1}^{q} \delta(y_i\!-\!a_k) \big(Q_{k\!-\!1}(p_i,\eta_i) \!-\!Q_{k}(p_i,\eta_i)\big)}, 
\end{align*}
for $i = 1,\dots, M$, with
\begin{align}\label{eq:gout_1}
\begin{cases}
Q_k(p,\eta) = Q(\frac{c_k - p}{\sqrt{\eta}})\\
Q^{'}_k(p,\eta) = \frac{e^{-(c_k-p)^2/(2\eta)}}{\sqrt{2\pi\eta}}\\
Q^{''}_k(p,\eta) =  Q^{'}_k(p,\eta) \frac{c_k -p}{\eta},
\end{cases}
\end{align}
where $Q(\cdot)$ in the first equation of \eqref{eq:gout_1} denotes the standard Gaussian $Q$-function while the $c_k$'s are given in \eqref{eq:quantizer_qary_1}. Steps \ref{step:gin} and \ref{step:gin_p} depend on the prior $p_0$. The function $g_\text{in}$ of step \ref{step:gin} acts on a $B$-dimensional sections and it is defined as follows
\begin{align*}
	[g_{\text{in}}({\br}_l,\boldsymbol{\tau}_l)]_i = 
	\Big[ 1+ \sum_{j\neq i}^B \exp\Big(\frac{2{r}_{lj}-1}{2 \tau_{lj}} -  \frac{2{r}_{li}-1}{2 \tau_{li}} \Big)\Big]^{-1},
\end{align*}
for $i = 1,\dots, B$, where ${\br}_l$ is the $l^{th}$ section ($l \in \{1,\dots,L\}$) and ${r}_{li}$ is the $i^{th}$ component of that section.

The GAMP algorithm requires an exchange of $\mathcal{O}(N)$ messages. The complexity of computing each message is dominated by a matrix-vector multiplication. In fact, both the matcher and the GAMP dematcher involve matrix-vector multiplication. Hence, the worst case complexity, per message, is $\mathcal{O}(MN)$.
This can be simplified using structured operators such as Fourier, wavelet or Hadamard. While random Gaussian matrices are mathematically more tractable and easier for analysis, the structured matrices provide practical advantages and more robust finite-length performance \cite{BarbierSchulkeKrzakala_15}. Hadamard-based matrices constructed as in \cite{BarbierSchulkeKrzakala_15}, with random sub-sampled modes of the full Hadamard operator, allow to achieve a complexity of $\mathcal{O}\big(m\ln(N)\big)$ and drastically reduce the storage need. Note that using such matrices might necessitate fine tuning the quantizer \eqref{eq:quantizer} as the codeword's distribution deviates from Gaussian. Moreover, one would need to use other variants of AMP that are better suited to general matrices.

\section{Performance Evaluation}
An important aspect of GAMP algorithm is that its asymptotic performance can be analytically tracked at each iteration by the state evolution (SE) equation.
SE is a simple recursion analogous to the density evolution used to track the performance of low-density parity-check (LDPC) codes on sparse graphical models. Moreover, the ultimate Bayes-optimal performance of our proposed scheme, i.e. the performance under optimal algorithm, can be obtained from the potential function \cite{Barbier_17_capacity-achieving,Dia_17_transactions} inspired from statistical physics techniques. Note that the GAMP performance is typically assessed using the mean-square error (MSE) between $\bs$ and $\hat{\bs}$ or the section error rate (SER) between $\bs$ and $\bar{\bs}$ (i.e. the fraction of sections that are wrongly reconstructed, see Fig. \ref{fig:blockDiag} for $\hat{\bs}$ and $\bar{\bs}$).

\begin{figure}[!t]
\centering
\includegraphics[draft=false,width=.2412\textwidth, height = 110pt]{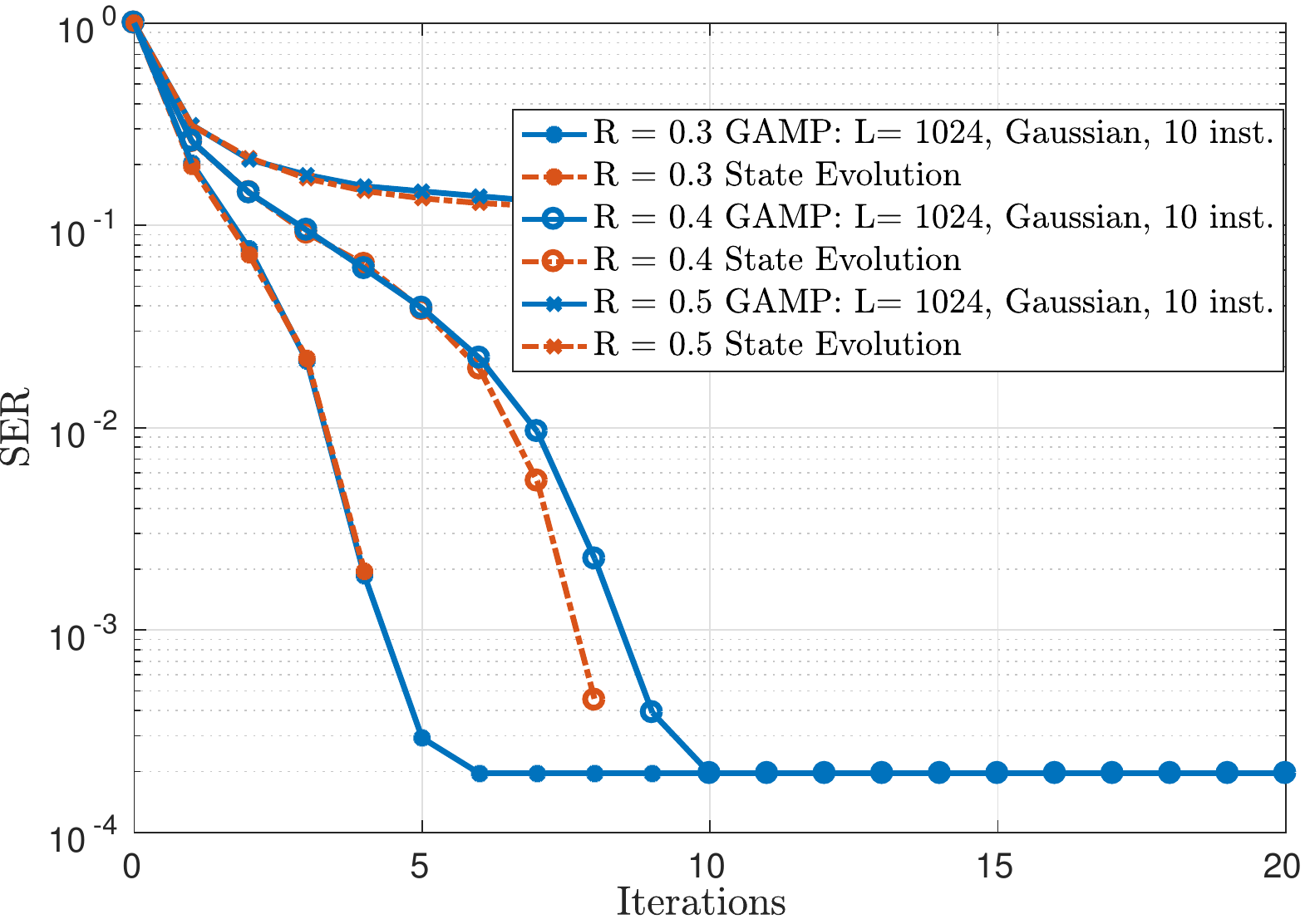}
\includegraphics[draft=false,width=.2412\textwidth, height = 110pt]{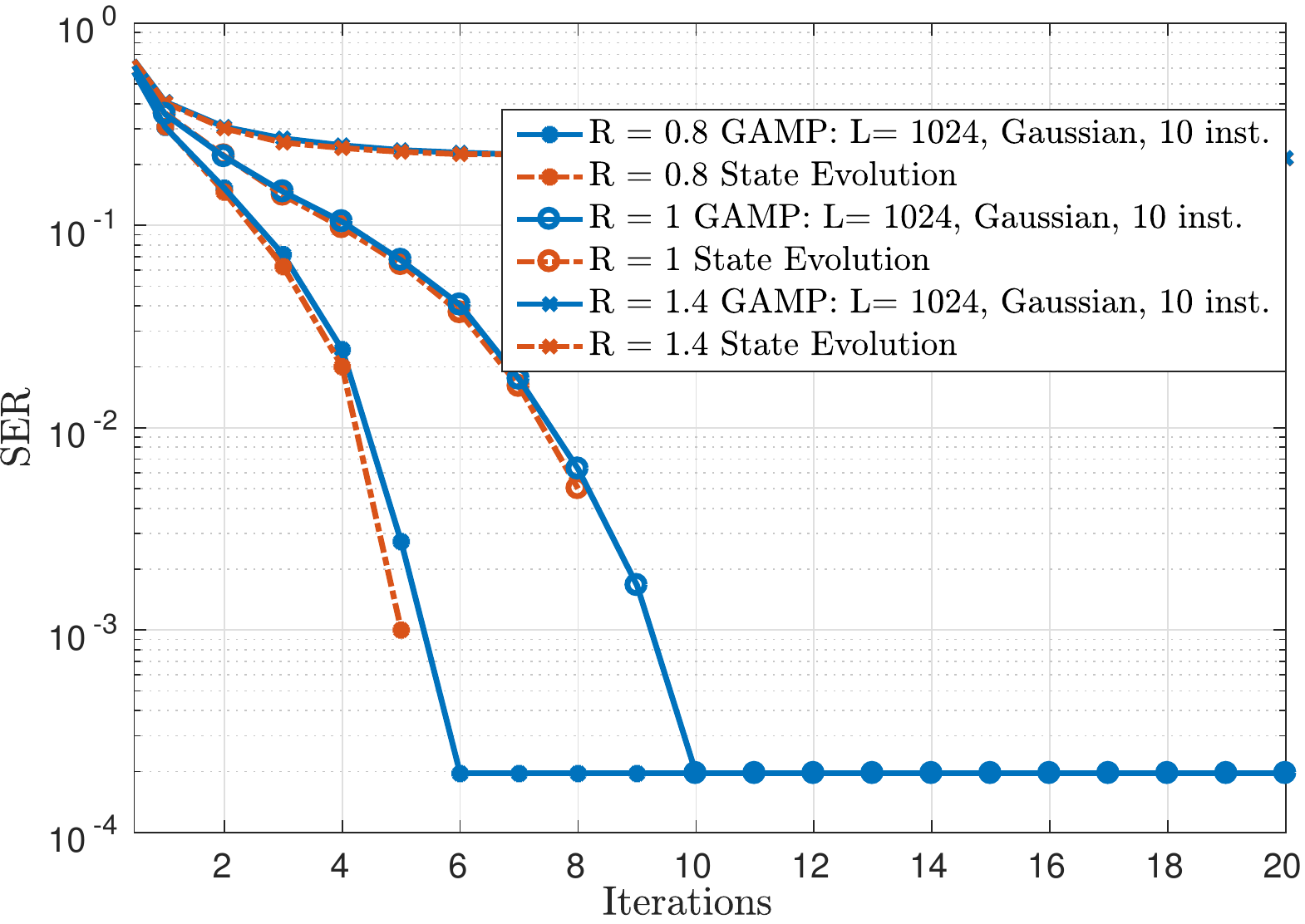}
\caption{The performance of GAMP at each iteration in terms of SER under Gaussian coding matrices. Left: binary-to-binary distribution matching with target $p^{\star}=1/4$; the information theoretical rate of such scheme is $R = h_2(p^{\star}) = 0.8113$. Right: binary to $4$-ary distribution matching with target distribution $P_{Y} = [0.1, 0.4, 0.3,0.2]$; the information theoretical rate of such scheme is $R = H(Y) = 1.8464$. We fix $B=4$ and we simulate GAMP under various rates. As long as the rate is small enough (i.e. $R<R_{\rm GAMP}$), GAMP (solid blue line) performs the dematching task up to a finite-length error floor that vanishes with $L$.
As $L$ increases, the GAMP performance coincides with the SE prediction (dotted red line) and the error floor vanishes.}
\label{fig:asymptotic}
\end{figure}

\begin{figure}[!t]
\centering
\includegraphics[draft=false,width=.42\textwidth, height = 140pt]{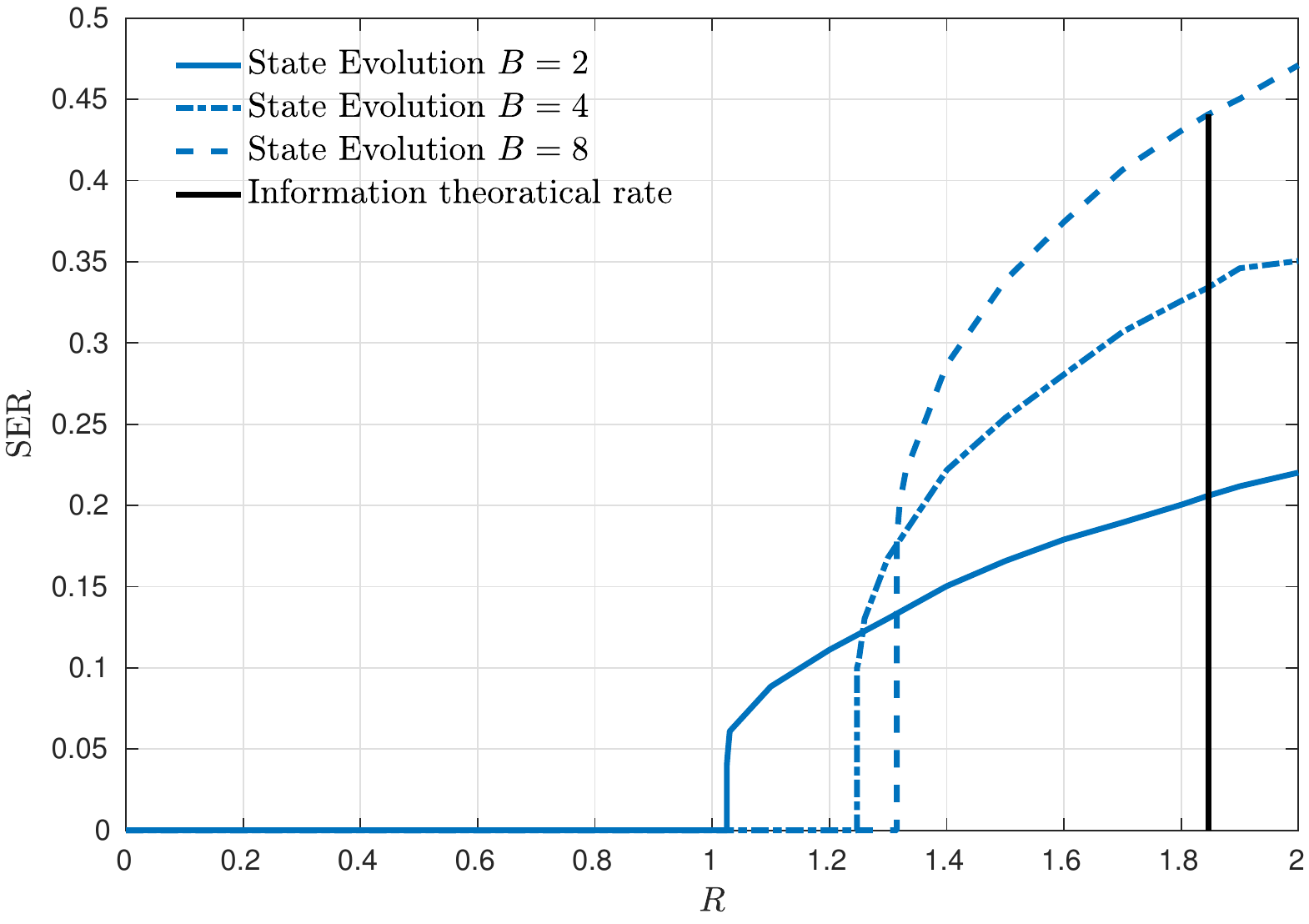}
\caption{The performance of GAMP as predicted by the SE for binary to $4$-ary distribution matching with target distribution $P_{Y} = [0.1, 0.4, 0.3,0.2]$.
We perform the SE analysis for three different section sizes. A sharp phase transition occurs for each section size: $R_{\rm GAMP} = 1.025$ for $B=2$,  $R_{\rm GAMP} = 1.247$ for $B=4$,  $R_{\rm GAMP} = 1.315$ for $B=8$. Note that although the gap to the information theoretical rate varies with $B$, a nonnegligible gap persists with the current construction as $B$ increases.}
\label{fig:phaseTransition}
\end{figure}

Numerical simulations as well as SE analysis show the following: for any fixed section size $B$, the GAMP algorithm exhibits asymptotically in $L$ a ``phase transition'' at an algorithmic rate (or threshold) denoted by $R_{\rm GAMP}$. Formally, $R_{\rm GAMP}$ is the maximum rate at which the GAMP algorithm performs the dematching task with vanishing reconstruction error.
These observations are depicted in Fig. \ref{fig:asymptotic} for both binary-to-binary and binary to $q$-ary distribution matching with $B=4$.
Our empirical results shown in Fig. \ref{fig:asymptotic} confirm that the SE tracks the asymptotic performance of GAMP dematcher. An alternative way to see the phase transition behavior is presented in Fig. \ref{fig:phaseTransition} where the final SER, after SE convergence, is plotted as a function of the rate for three different section sizes.
%

\begin{figure}[!t]
\centering
\includegraphics[draft=false,width=.42\textwidth, height = 140pt]{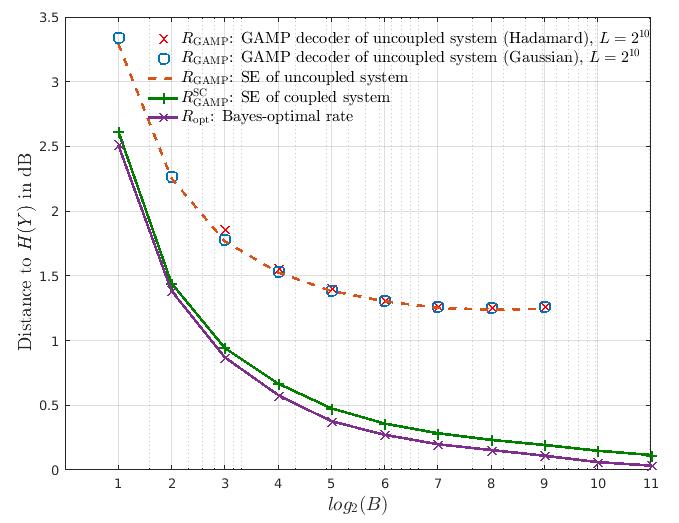}
\caption{The distance between the algorithmic rate and $H(Y)$ in dB as a function of the section size. Binary-to-binary distribution matching is performed with target $p^{\star}=1/4$. For the uncoupled system: the gap to the information theoretical rate persists even with large section sizes (dotted line). The performance of GAMP under Gaussian matrices (circles) is accurately predicted by the SE analysis. Applying GAMP
on Hadamard-based matrices (crosses) yields a small mismatch w.r.t. the SE prediction. This lack of SE accuracy for non-Gaussian matrices can be handled by using other variants of AMP. 
For the spatially coupled system: the algorithmic rate (green curve) follows the optimal rate (purple curve) that tends to $H(Y)$. Spatial coupling is performed with the following coupling parameters: $L_c=32$, $L_r$ = 33, $w_b=2$, $w_f=2$, $\beta=1.2$ and $J=0.1$. The small mismatch between the purple and green curves is due to the finite length of coupling parameters. As the coupling parameters increase, the two curves coincide.
}
\label{fig:GaussianHadamard}
\end{figure}

The empirical algorithmic rate $R_{\rm GAMP}$, obtained from running GAMP on real instances, as well as the one obtained from the numerical SE analysis for the current ``uncoupled'' construction are shown on the upper curve of Fig. \ref{fig:GaussianHadamard} for different values of $B$. Under Gaussian coding matrices, the performance is accurately predicted by the SE for all values of $B$. Using Hadamard-based matrices incurs a small performance loss, in terms of $R_{\rm GAMP}$, that vanishes with $B$. However, a gap to the information theoretical rate persists as  $B$ increases.

The gap to the information theoretical rate is due to the sub-optimality of GAMP, which is a low-complexity iterative algorithm. In order to predict the Bayes-optimal performance of our proposed scheme under optimal algorithm, which is computationally intractable, we use the potential function. Numerical simulations show that the Bayes-optimal rate (or potential threshold) denoted by $R_{\rm opt}$ approaches the information theoretical rate as the section size increases (see Fig. \ref{fig:GaussianHadamard}). Moreover, using a heuristic analysis of the potential function as in \cite{Barbier_17_capacity-achieving,Dia_17_transactions}, one can argue that $R_{\rm opt}$ indeed tends to the information theoretical rate as $B \rightarrow \infty$.
\begin{figure}[!t]
\centering
\includegraphics[draft=false,width=.42 \textwidth, height = 100pt]{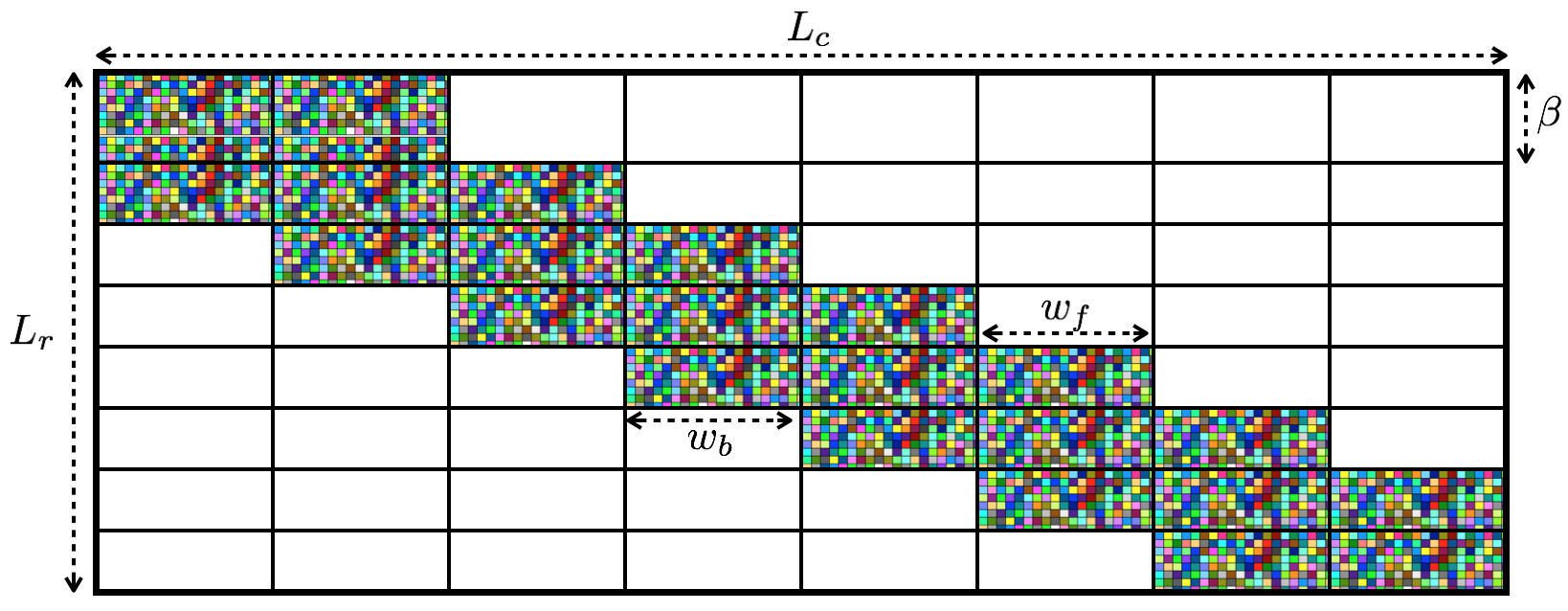}
\caption{A spatially coupled coding matrix $\bF$ with $L_r \times L_c$ blocks. Besides the diagonal blocks, there are $w_b$ and $w_f$ off-diagonal blocks with nonzero elements representing the backward and forward coupling windows respectively. Each block is of dimension $M \times N$ except the blocks in the first block-row that have a dimension of $\beta M \times N$, where $\beta$ represents the seed rate. Each nonzero block is composed of i.i.d. real Gaussian entries with zero mean and a certain variance such that the variances of each block-row add up to 1. The variances can be tuned in a uniform or nonuniform fashion using the coupling strength parameter $J$ \cite{Barbier_17_capacity-achieving,BarbierSchulkeKrzakala_15}.}
\label{fig:spatialCoupledMatrix}
\end{figure}

An effective approach to boost the algorithmic performance of GAMP is to apply spatial coupling as done for capacity achieving SS codes \cite{Barbier_17_capacity-achieving,BarbierSchulkeKrzakala_15} (see Fig. \ref{fig:spatialCoupledMatrix} for the construction of spatially coupled coding matrix). Spatial coupling significantly improves the performance and decreases the gap to 
$H (Y)$. The gap can be made arbitrarily small by increasing $B$ (see Fig. \ref{fig:GaussianHadamard}). Our SE simulations show that the algorithmic rate of the spatially coupled system denoted by $R_{\rm GAMP}^{\rm SC}$ follows the Bayes-optimal rate, which tends to the information theoretical rate. Actually, one can show that the algorithmic rate of the spatially coupled system equals the Bayes-optimal rate in a proper limit as done for SS codes \cite{Dia_16_ITW,Dia_17_transactions}. This phenomenon turns to be quite general and it is coined \emph{threshold saturation}.
%
%
\section{Conclusion}
In this work, we present a novel formulation of the fixed-length distribution matching inspired from the SS codes and the compressed sensing paradigm. The proposed solution uses a low-complexity dematching based on the GAMP algorithm. We show that GAMP dematching along with spatial coupling yields asymptotically optimal performance. Moreover, we investigate practical scenarios using Hadamard-based operators. A notable aspect of the proposed solution is the amenability to perform joint channel coding and matching.    
\bibliographystyle{IEEEtran}
\bibliography{IEEEabrv,bibliography}

\begin{thebibliography}{10}
\providecommand{\url}[1]{#1}
\csname url@samestyle\endcsname
\providecommand{\newblock}{\relax}
\providecommand{\bibinfo}[2]{#2}
\providecommand{\BIBentrySTDinterwordspacing}{\spaceskip=0pt\relax}
\providecommand{\BIBentryALTinterwordstretchfactor}{4}
\providecommand{\BIBentryALTinterwordspacing}{\spaceskip=\fontdimen2\font plus
\BIBentryALTinterwordstretchfactor\fontdimen3\font minus
  \fontdimen4\font\relax}
\providecommand{\BIBforeignlanguage}[2]{{%
\expandafter\ifx\csname l@#1\endcsname\relax
\typeout{** WARNING: IEEEtran.bst: No hyphenation pattern has been}%
\typeout{** loaded for the language `#1'. Using the pattern for}%
\typeout{** the default language instead.}%
\else
\language=\csname l@#1\endcsname
\fi
#2}}
\providecommand{\BIBdecl}{\relax}
\BIBdecl

\bibitem{Forney_84}
G.~Forney, R.~Gallager, G.~Lang, F.~Longstaff, and S.~Qureshi, ``Efficient
  modulation for band-limited channels,'' \emph{IEEE Journal on Selected Areas
  in Communications}, vol.~2, no.~5, pp. 632--647, Sep 1984.

\bibitem{Steiner_Bocherer_16_comparison}
F.~Steiner and G.~Böcherer, ``Comparison of geometric and probabilistic
  shaping with application to atsc 3.0,'' in \emph{SCC 2017; 11th International
  ITG Conference on Systems, Communications and Coding}, Feb 2017, pp. 1--6.

\bibitem{Mondelli_Urbanke_Hassani_14_assymetricChannels}
M.~Mondelli, R.~Urbanke, and S.~H. Hassani, ``How to achieve the capacity of
  asymmetric channels,'' in \emph{Communication, Control, and Computing
  (Allerton), 2014 52nd Annual Allerton Conference on}, Sep 2014, pp. 789--796.

\bibitem{MacKay_99_good-error-correcting}
D.~J.~C. MacKay, ``Good error-correcting codes based on very sparse matrices,''
  \emph{IEEE Transactions on Information Theory}, vol.~45, no.~2, pp. 399--431,
  Mar 1999.

\bibitem{Kschischang_Pasupathy_93_optimal-nonuniform}
F.~R. Kschischang and S.~Pasupathy, ``Optimal nonuniform signaling for gaussian
  channels,'' \emph{IEEE Transactions on Information Theory}, vol.~39, no.~3,
  pp. 913--929, May 1993.

\bibitem{Ungerboeck_02_huffman-shaping}
G.~Ungerboeck, \emph{Huffman Shaping}.\hskip 1em plus 0.5em minus 0.4em\relax
  Boston, MA: Springer US, 2002, pp. 299--313.

\bibitem{Bocherer_11_dyadic}
G.~Böcherer and R.~Mathar, ``Matching dyadic distributions to channels,'' in
  \emph{Data Compression Conference (DCC 2011)}, Snowbird, USA, Mar. 2011, pp.
  23--32.

\bibitem{Amjad_Bocherer_13_fixed-to-variable}
R.~A. Amjad and G.~Böcherer, ``Fixed-to-variable length distribution
  matching,'' in \emph{2013 IEEE International Symposium on Information
  Theory}, July 2013, pp. 1511--1515.

\bibitem{Cai_Ho_Yeung_07_probabilistic-capacity}
N.~Cai, S.~W. Ho, and R.~W. Yeung, ``Probabilistic capacity and optimal coding
  for asynchronous channel,'' in \emph{2007 IEEE Information Theory Workshop},
  Sept 2007, pp. 54--59.

\bibitem{Baur_Bocherer_15_arithmetic}
N.~Baur and G.~Böcherer, ``Arithmetic distribution matching,'' in \emph{SCC
  2015; 10th International ITG Conference on Systems, Communications and
  Coding}, Feb 2015, pp. 1--6.

\bibitem{Schulte_Bocherer_16_constant}
P.~Schulte and G.~Böcherer, ``Constant composition distribution matching,''
  \emph{IEEE Transactions on Information Theory}, vol.~62, no.~1, pp. 430--434,
  Jan 2016.

\bibitem{Donoho_06_compressed-sensing}
D.~L. Donoho, ``Compressed sensing,'' \emph{IEEE Transactions on Information
  Theory}, vol.~52, no.~4, pp. 1289--1306, April 2006.

\bibitem{Candes_Tao_06_near-optimal}
E.~J. Candes and T.~Tao, ``Near-optimal signal recovery from random
  projections: Universal encoding strategies?'' \emph{IEEE Transactions on
  Information Theory}, vol.~52, no.~12, pp. 5406--5425, Dec 2006.

\bibitem{Barron_Joseph_10_toward-fast-reliable}
A.~R. Barron and A.~Joseph, ``Toward fast reliable communication at rates near
  capacity with gaussian noise,'' in \emph{2010 IEEE International Symposium on
  Information Theory}, June 2010, pp. 315--319.

\bibitem{Barbier_17_capacity-achieving}
J.~Barbier and F.~Krzakala, ``Approximate message-passing decoder and capacity
  achieving sparse superposition codes,'' \emph{IEEE Transactions on
  Information Theory}, vol.~63, no.~8, pp. 4894--4927, Aug 2017.

\bibitem{Rush_17}
C.~Rush, A.~Greig, and R.~Venkataramanan, ``Capacity-achieving sparse
  superposition codes via approximate message passing decoding,'' \emph{IEEE
  Transactions on Information Theory}, vol.~63, no.~3, pp. 1476--1500, March
  2017.

\bibitem{Dia_16_ISIT}
J.~Barbier, M.~Dia, and N.~Macris, ``Proof of threshold saturation for
  spatially coupled sparse superposition codes,'' in \emph{2016 IEEE
  International Symposium on Information Theory (ISIT)}, July 2016, pp.
  1173--1177.

\bibitem{Donoho_09}
D.~L. Donoho, A.~Maleki, and A.~Montanari, ``Message-passing algorithms for
  compressed sensing,'' \emph{Proceedings of the National Academy of Sciences},
  vol. 106, no.~45, pp. 18\,914--18\,919, 2009.

\bibitem{Rangan_12_gamp}
S.~Rangan, ``Generalized approximate message passing for estimation with random
  linear mixing,'' \emph{arXiv preprint arXiv:1010.5141}, 2012.

\bibitem{QunatizedCS_2012_Rangan}
U.~S. Kamilov, V.~K. Goyal, and S.~Rangan, ``Message-passing de-quantization
  with applications to compressed sensing,'' \emph{IEEE Transactions on Signal
  Processing}, vol.~60, no.~12, pp. 6270--6281, Dec 2012.

\bibitem{Dia_16_ITW}
J.~Barbier, M.~Dia, and N.~Macris, ``Threshold saturation of spatially coupled
  sparse superposition codes for all memoryless channels,'' in \emph{2016 IEEE
  Information Theory Workshop (ITW)}, Sept 2016, pp. 76--80.

\bibitem{Dia_17_ISIT}
E.~Biyik, J.~Barbier, and M.~Dia, ``Generalized approximate message-passing
  decoder for universal sparse superposition codes,'' in \emph{2017 IEEE
  International Symposium on Information Theory (ISIT)}, June 2017, pp.
  1593--1597.

\bibitem{Dia_17_transactions}
J.~Barbier, M.~Dia, and N.~Macris, ``Universal sparse superposition codes with
  spatial coupling and gamp decoding,'' \emph{arXiv preprint arXiv:1707.04203},
  2017.

\bibitem{Caire_03_lossless}
G.~Caire, S.~Shamai, and S.~Verdu, ``Lossless data compression with error
  correcting codes,'' in \emph{2003 IEEE International Symposium on Information
  Theory (ISIT)}, June 2013, pp. 22--26.

\bibitem{Cronie_10_losslessPolar}
H.~S. Cronie and S.~B. Korada, ``Lossless source coding with polar codes,'' in
  \emph{2010 IEEE International Symposium on Information Theory}, June 2010,
  pp. 904--908.

\bibitem{BarbierSchulkeKrzakala_15}
J.~Barbier, C.~Sch\"{u}lke, and F.~Krzakala, ``Approximate message-passing with
  spatially coupled structured operators, with applications to compressed
  sensing and sparse superposition codes,'' \emph{Journal of Statistical
  Mechanics: Theory and Experiment}, vol. 2015, no.~5, p. P05013, 2015.

\end{thebibliography}

\end{document}